\documentclass[amssymb]{revtex4}
\usepackage{float}
\usepackage{epsfig,graphicx}
\usepackage{amsmath}
\usepackage{amsfonts}
\usepackage{amssymb}

\newcommand{\be}{\begin{equation}}
\newcommand{\ee}{\end{equation}}
\newcommand{\beq}{\begin{equation}}
\newcommand{\eeq}{\end{equation}}
\newcommand{\beqa}{\begin{eqnarray}}
\newcommand{\eeqa}{\end{eqnarray}}
\newcommand{\bea}{\begin{eqnarray}}
\newcommand{\eea}{\end{eqnarray}}

\newcommand{\om}{\omega}

\newcommand{\D}{\partial}
\newcommand{\zb}{\bar{z}}

\newcommand{\rmd}{\mathrm{d}}
\newcommand{\rme}{\mathrm{e}}
\newcommand{\rmi}{\mathrm{i}}
\newcommand{\rmt}{\mathrm{t}}
\newcommand{\rmc}{\mathrm{c}}
\newcommand{\mh}{\frac{m}{2}}

\begin{document}
\title
{\large \bf Exact results for the spectra of interacting bosons and fermions \\
on the lowest Landau level}

\author{ Stefan Mashkevich\footnote{mash@mashke.org}}
\affiliation{Schr\"odinger, 120 West 45th St., New York, NY 10036, USA}
\affiliation{
Bogolyubov Institute for Theoretical Physics, Kiev 03143, Ukraine}
\author{ Sergey Matveenko\footnote{matveen@landau.ac.ru}}
\affiliation{Landau Institute for Theoretical Physics, Kosygina Str. 2,
119334, Moscow, Russia}
\author{ St\'ephane Ouvry\footnote{stephane.ouvry@u-psud.fr}}
\affiliation{Laboratoire de Physique Th\'eorique et Mod\`eles
Statistiques\footnote{{\it Unit\'e de
Recherche de l'Universit\'e Paris 11 associ\'ee au CNRS, UMR 8626}}\\
B\^at. 100, Universit\'e Paris-Sud, 91405 Orsay, France}

\date{\today}
\begin{abstract}
A system of $N$ interacting bosons or fermions in a two-dimensional harmonic potential
(or, equivalently, magnetic field) whose states are projected onto the
lowest Landau level is considered.
Generic expressions are derived for matrix elements of any interaction,
in the basis of angular momentum eigenstates.
For the fermion ``ground state'' ($N=1$ Laughlin state), this makes it possible to exactly
calculate its energy all the way up to the mesoscopic regime $N \sim 1000$.
It is also shown that for $N = 3$ and Coulomb interaction,
several rational low-lying values of energy exist, for bosons and fermions alike.
\end{abstract}

\maketitle

\section{Introduction}

This paper is a sequel of Ref.~\cite{LLL}, where exact eigenstates were discussed for bosons with
contact interaction in the lowest Landau level (LLL) of a strong magnetic field
in two dimensions, as well as eigenenergies  for fermions with
Laplacian delta interaction, for which Laughlin wavefunctions are known to be exact eigenstates.
In this paper, general expressions for matrix elements of an arbitrary central
interaction --- a sum of two body-interactions $V(r_{ij})$ whose Fourier transform
admits a Laurent expansion $v(k)={a_{-1}\over k}+a_0+a_1k+\ldots$ ---
projected onto the lowest Landau level are derived.
These include $1/r^n$ with any $n$, as well as contact (delta) and Laplacian delta interactions.
An exact expression of the interaction energy $E(N)$ for the $N$-fermion
``ground state'' (the $n=1$ Laughlin state, which is actually the ground state
in the presence of a harmonic potential)
is derived from which the large $N$ asymptotic behavior can be obtained.
For Coulomb interactions, the asymptotics is $E(N)\propto N^{3/2}$,
which is confirmed by direct numerical calculation  up to $N = 1000$.
Also for Coulomb interactions, in the three-body problem, rational values of energy exist
for low values of the total angular momentum, for bosons and fermions alike.

Clearly, on the experimental side, we have in mind rotating Bose-Einstein condensates \cite{BEC}
on the one hand, and strongly correlated Quantum Hall fermion droplets \cite{QHFD}
on the other hand. In both cases a magnetic field is present,
be it real in the quantum Hall case, or effective (due to the rotation of the condensate) in the BEC case.
In the sequel, as a matter of simplification, we consider a harmonic trap one-body Hamiltonian,
and the projection of the interaction is  made on the one-body  harmonic eigenstates
\be \label{66}
\langle z,\bar z|0,l\rangle = \left({{\omega ^{l+1}\over \pi l!}}\right)^{{1\over 2}}z^{l}
\rme^{-{1\over 2}\omega z\bar z}\;,
\ee
Indeed, if a magnetic field were added to the harmonic trap, the one-body eigenstates
corresponding to the LLL (Landau level number $n=0$, angular momentum $l\ge 0$)
would be the LLL-harmonic eigenstates basis (in complex coordinates)
\be \label{6}
\langle z,\bar z|0,l\rangle = \left({{\omega_\rmt ^{l+1}\over \pi l!}}\right)^{{1\over 2}}z^{l}
\rme^{-{1\over 2}\omega_\rmt z\bar z}\to_{\omega_\rmc\to 0} \left({{\omega ^{l+1}\over \pi l!}}\right)^{{1\over 2}}z^{l}
\rme^{-{1\over 2}\omega z\bar z}\;,
\ee
where $\omega_\rmt = \sqrt{\omega_\rmc^2 + \omega^2}$, $\omega_\rmc$ being
half the cyclotron frequency. Note that since we diagonalize the system
in a given angular momentum sector (the angular momentum operator commutes
with the interaction Hamiltonian), the magnetic field simply shifts
the total energy by a constant term, which is therefore ignored here.

\section{MATRIX ELEMENTS}

The Hamiltonian for $N$ interacting particles in a harmonic trap is
\beq
{\cal{H}} = -2\sum_{i=1}^N \D_i \bar\D_i + \frac{\omega^2}{2} \sum_{i=1}^N z_i\zb_i
+ \sum_{i<j=1}^N V(|z_i - z_j|) \;.
\label{H}
\eeq
As long as the interaction potential $V(|z_i - z_j|)$ vanishes at infinity,
so that the harmonic potential dominates,
the asymptotics of the wave function can be detached as usually,
\beq
 \psi(z_1, \zb_1, ...,z_N, \zb_N)=\exp\left(-\frac{\omega}{2} \sum_{i=1}^N
z_i\zb_i\right)\chi(z_1, \zb_1, ...,z_N, \zb_N) \;;
\eeq
then the Hamiltonian acting on $\chi$ is
\beq
H = H_0 + \sum_{i<j=1}^N V(|z_i - z_j|)
\eeq
where the free Hamiltonian $H_0$ is
\beq
H_0 =  \sum_{i=1}^N (-2 \D_i \bar\D_i + \om + \om z_i \D_i + \om \zb_i \bar\D_i ) \;.
\eeq
%and the interaction is unchanged:
%\beq
%V = \sum_{i<j=1}^N V(|z_i - z_j|) \;.
%\eeq

From now on one sets $\omega=1$. The LLL projector has the form
\beq
P \chi  =  \prod_{i=1}^N
\left[\frac{1}{\pi}\int \rme^{-z'_i \zb'_i + z_i \zb'_i}  \rmd z'_i \rmd \zb'_i \right]
\chi (z'_1, \zb'_1,...,z'_N, \zb'_N) \;.
\label{LLLP}
\eeq
LLL functions are analytic
\beq
\chi = \chi(z_1, ... ,z_N) \;;
\eeq
for such a function, $P \chi = \chi$, and, since $\chi$ does not depend
on $\zb$'s,
\beq
P H_0 \chi  = \om(N + \sum_{i=1}^N z_i \D_i) \chi \;.
\eeq

Perform a Fourier transform of the interaction,
\beq
V(r) = \int \frac{\rmd^2 { \vec k}}{2 \pi }\rme^{\displaystyle\rmi {\vec  k \vec r}} v(k)\;.
\eeq
and introduce the complex coordinates
%\beq
%\frac{1}{r} = \int \frac{d^2 {\bf k}}{2 \pi }\frac{\rme^{i {\bf k r}}}{k},
%\eeq
\beq
{\vec k \vec r} = \frac{{\bf \bar  k} z}{2} + \frac{{\bf k} \zb}{2}, \quad
{\bf k} = k_x +\rmi k_y \;.
\eeq
One finally  obtains the LLL-projected interaction
\beqa
P \sum_{i<j=1}^N V(|z_i - z_j|) \chi & = & \sum_{i<j=1}^N \int \exp\left[\rmi\frac{ \bar {\bf k}(z_i - z_j)}{2}\right]
\frac{v(k)}{2\pi} \,
\chi( ...\,, z_i + \rmi\frac{ {\bf k}}{2}, ... \,, z_j - \rmi\frac{ {\bf k}}{2}, ...) \,
\rme^{-\frac{k^2}{2}} \rmd {\bf k} \, \rmd{\bf{\bar k}}\nonumber \\
& = & \sum_{i<j=1}^N \sum_{n_1, n_2 = 0}^\infty
\int_0^{\infty} k^{2(n_1+n_2)+1} v(k) \rme^{-\frac{k^2}{2}} \rmd k
\frac{(-1)^{n_1} (z_i -z_j)^{n_1+n_2}}{n_1!\,n_2!\,(n_1+n_2)!\,2^{2(n_1+n_2)}}
\frac{\D^{n_1+n_2}}{\D z_i^{n_1}\D z_j^{n_2}} \chi(z_1, ...,z_N) \nonumber \\
& = & \sum_{i<j=1}^N \sum_{n=0}^\infty
\int_0^{\infty} k^{2n+1} v(k) \rme^{-\frac{k^2}{2}} \rmd k
\frac{ (z_i -z_j)^{n}(\D_j -\D_i)^n}{(n!)^2\,2^{2n}}\chi(z_1, ...,z_N)\;,
\label{pV}
\eeqa
where one has expanded every term containing the $z$'s
into Taylor series, substituted ${\bf k} = k \exp( \rmi \phi)$,
and integrated over $\phi$.

In the last expression of (\ref{pV}) the expansion coefficient $\int_0^{\infty} k^{2n+1} v(k) \rme^{-\frac{k^2}{2}} \rmd k $ is well defined as long as $v(k)$ admits a Laurent expansion $v(k)=\sum_{m=-1}^{\infty}a_m k^m$ such that

\be \label{gamma}\int_0^{\infty} k^{2n+1} k^m  \rme^{-\frac{k^2}{2}} \rmd k = 2^{ n+{m\over 2}}\Gamma (n+1+{m\over 2})\ee

An elementary LLL $N$-body wave function is
\beq
\chi = \prod_{i=1}^N z_i^{l_i} \;;
\label{basis}
\eeq
for bosons (fermions), it has to be (anti)symmetrized.
It is an eigenfunction of the total angular momentum,
 $L=\sum_{i=1}^N l_i$, and thus an eigenfunction of $H_0$.
It is obvious from (\ref{pV}) that $P\sum_{i<j=1}^N V(|z_i - z_j|)$ conserves the angular momentum
(as any central interaction should do), therefore it is enough
to diagonalize it in each sector of given $L$.

The states with the lowest absolute value of angular momentum
--- for brevity, we will refer to them as ``ground states''
(which they are if there is a harmonic potential), both of bosons,
\beq
\chi_{0}^\mathrm{B}(N) = 1\;, \quad L = 0 \;,
\eeq
and of fermions,
\beq
\chi_{0}^\mathrm{F}(N) = \prod_{i<j=1}^N (z_i - z_j)\;, \quad L = \frac{N(N-1)}{2} \;,
\eeq
are nondegenerate with respect to angular momentum,
which implies that they are both eigenfunctions of
$P\sum_{i<j=1}^N V(|z_i - z_j|)$.
%(When we act on the ground state with the interacting potential and then project with $P$,
%anything that does not belong to the LLL, the only surviving state  is the ground state itself.)
For bosons, this can be seen directly by looking at Eq.~(\ref{pV})
where only the $n= 0$ term survives so that
\beq
P\sum_{i<j=1}^N V(|z_i - z_j|)  = \frac{N(N-1)}{2} \int v(k) k \, \rme^{-\frac{k^2}{2}} \rmd k \;
\eeq
and thus $$E(N)=\frac{N(N-1)}{2} \int v(k) k \, \rme^{-\frac{k^2}{2}} \rmd k$$
(the diagonal matrix element of the interaction in the momentum representation.)
%For fermions, it is not so obvious. We have verified by a direct calculation
%that it works for Coulomb interaction where $v(k) \propto 1/k$, for $N=3$ particles.
%A~general proof might involve expanding $v(k)$ into a Laurent series and showing
%that it works for each separate power in that expansion,
%when the integral in Eq.~(\ref{pV}) reduces to the gamma function (\ref{gamma}).

\section{Fermion ground state}
In the Fermi case it is also possible to solve
the eigenvalue equation
\be
P\sum_{i<j=1}^N V(|z_i - z_j|)\chi_0^\mathrm{F}(N) = E(N)\chi_0^\mathrm{F}(N) \;,
\label{PVeigen}
\ee
where $P\sum_{i<j=1}^N V(|z_i - z_j|)$  is given in (\ref{pV}).

The key observation is that, since said state $\chi_0^\mathrm{F}(N)$ is known to be
an eigenstate of $P\sum_{i<j=1}^N V(|z_i - z_j|)$
it is not necessary to calculate the whole LHS of (\ref{PVeigen}).
Being a Vandermonde determinant $\chi_0^\mathrm{F}(N)$ rewrites as
% --- it suffices to calculate the coefficient in front
%of a {\em single} term of
%One has
\be
\chi_{0}^\mathrm{F}(N) = \prod_{l=1}^N z_l^{l-1} +  \cdots\;,
\ee
where the omitted $(N!-1)$ terms come from antisymmetrization. It follows that the coefficient in front of $\prod_{l=1}^N z_l^{l-1}$
on the LHS of Eq.~(\ref{PVeigen}) is necesseraly $E(N)$.
Return to the second line of Eq.~(\ref{pV}) and let $\chi(z_1, ...,z_N)$ be the monomial
$\prod_{l=1}^N z_l^{p_l}$.
Then a term with given $i$ and $j$ in the sum on the RHS of that equation
will be a sum of monomials in each of which only the powers of $z_i$ and $z_j$
are different from $p_i$ and $p_j$, respectively
(and the power of any $z_l$ with $l \ne i,j$ is still $p_l$).
Moreover, the sum of all powers of $z$'s, which is the total angular momentum, never changes.
Hence, there are only two cases when $P\sum_{i<j=1}^N V(|z_i - z_j|)\chi$
can contain $\prod_{l=1}^N z_l^{l-1}$ as one of its terms:
%(i) $\chi = \prod_{i=1}^N z_i^{i-1}$; (ii) $\chi = z_1^0z_2^1 \cdots z_{i-1}^{i-2} z_{i}^{j-1} z_{i+1}^{i}
%\cdots z_{j-1}^{j-2} z_{j}^{i-1} z_{j+1}^{j} \cdots z_{N}^{N-1}$
(i) $p_l = l-1$; (ii) $p_l = l-1$ ($l \ne i,j$); $p_i = j-1$; $p_j = i-1$
(i.e., $z_i$ and $z_j$ are interchanged; in the Vandermonde determinant,
the corresponding monomial comes with a minus sign).
Moreover, in each of these two cases, for given $n_1$ and $n_2$, no more than
a single term in the binomial expansion of $(z_i -z_j)^{n_1+n_2}$ will yield the desired contribution.
In case (i), that term is $z_i^{n_1}z_j^{n_2}$ (so that the powers of $z_1$ and $z_2$ stay
unchanged after differentiation followed by multiplication);
in case (ii), it is $z_i^{n_1+i-j}z_j^{n_2+j-i}$ [so that the power of $z_i$, which is $j-1$ in $\chi$,
becomes $j-1-n_1+n_1+i-j = i-1$ in the $P\sum_{i<j=1}^N V(|z_i - z_j|)\chi$; likewise for $z_j$].
The maximum possible values of $n_1$ and $n_2$ are the powers of $z_i$ and $z_j$, respectively, in $\chi$.
Taking this into account and gathering all the coefficients, we obtain
\be
E(N) = \int_0^\infty f(N,k) v(k) k \, \rme^{-\frac{k^2}{2}} \rmd k \;,
\label{EN}
\ee
where
\be
f(N,k) = \sum_{i<j=1}^N \left[ \sum_{n_1=0}^i \sum_{n_2=0}^j c_{ijn_1n_2}(k)
- \sum_{n_1=0}^j \sum_{n_2=0}^i d_{ijn_1n_2}(k) \right]
\label{fNk}
\ee
with
\bea
c_{ijn_1n_2}(k) & = & \frac{(-1)^{n_1+n_2}}{(n_1!n_2!)^2} (i-n_1)_{n_1}(j-n_2)_{n_2}
\left(\frac{k}{2}\right)^{2(n_1+n_2)} \;,
\nonumber \\
d_{ijn_1n_2}(k) & = & \frac{(-1)^{n_1+n_2+j-i}}{n_1!n_2!(n_1+i-j)!(n_2+j-i)!} (j-n_1)_{n_1}(i-n_2)_{n_2}
\left(\frac{k}{2}\right)^{2(n_1+n_2)} \;;
\eea
$(x)_n \equiv \frac{\Gamma(x+n)}{\Gamma(x)}$ is the Pochhammer symbol.

The summation over $n_1$ and $n_2$ can be performed explicitly, by noting that
\bea
c_{ijn_1n_2}(k) & = & f_{iin_1}(k) f_{jjn_2}(k) \;,
\nonumber \\
d_{ijn_1n_2}(k) & = & (-1)^{j-i} f_{ijn_1}(k) f_{jin_2}(k) \;,
\eea
where
\be
f_{ijn}(k) = \frac{(-1)^n}{n!(n+i-j)!} (j-n)_n \left(\frac{k}{2}\right)^{2n} \;,
\ee
and that
\be
\sum_{n=0}^j f_{ijn}(k) = \frac{(j-1)!}{(i-1)!} L_{j-1}^{i-j}\left(\frac{k^2}{4}\right) \;.
\ee
Hence,
\be
f(N,k) = \sum_{i<j=1}^N \left[ L_{i-1}\left(\frac{k^2}{4}\right) L_{j-1}\left(\frac{k^2}{4}\right)
- (-1)^{j-i} L_{j-1}^{i-j}\left(\frac{k^2}{4}\right) L_{i-1}^{j-i}\left(\frac{k^2}{4}\right)\right] \;.
\label{fNkL}
\ee

For example for the first values of $N$ one has
\bea
f(2,k) & = & -\frac{k^2}{2} + 1 \;, \nonumber \\
f(3,k) & = & -\frac{k^6}{64} + \frac{9k^4}{32} - \frac{9k^2}{4} + 3 \;, \nonumber \\
f(4,k) & = & -\frac{k^{10}}{6144} + \frac{25k^8}{3072} - \frac{k^6}{6} + \frac{23k^4}{16} - 6k^2 + 6 \;.
\eea
Substituting into Eq.~(\ref{EN}) gives, within the
LLL-projection approximation,  the energy of the $N$-fermion Vandermonde state $\chi_0^\mathrm{F}(N)$
for any central pairwise interaction.

Further simplification is possible. A recurrency relation is
\be
f(N+1,k) - f(N,k) =  L_{N-1}^1\left(\frac{k^2}{4}\right) L_{N}\left(\frac{k^2}{4}\right)
+\sum_{i=1}^{N} \left[(-1)^{N-i} L_{N}^{i-N-1}\left(\frac{k^2}{4}\right)
L_{i-1}^{N+1-i}\left(\frac{k^2}{4}\right)\right] \;,
\label{fNkrec}
\ee
where in the first term, $\sum_{p=0}^n L_p^a (x) = L_n^{a+1}(x)$ has been used.
This  is much more efficient than (\ref{fNkL}), as it requires  to compute  a single sum
for each subsequent $N$, instead of a double sum.

One wants to find an expression for the chemical potential 
\be
\mu(N) \equiv E(N+1) - E(N)
\label{mu}
\ee

%The energy can be obtained by  integrating  the  chemical potential
%\beq
%\mu(N-1) = E(N)-E(N-1) = \int_0^\infty (f(N,k) - f(N-1,k)) v(k) k \, \rme^{-\frac{k^2}{2}} \rmd k
%\label{mu}
%\eeq
%Using  the first sum in (\ref{fNkrec})
%becomes {\bf***} $L^1_{N-2} (x) L_{N-1} (x)$ where $x=k^2/4$.
Using
\beq
L_n^a(x) = \frac{1}{n!} e^x x^{-a} \frac{d^n}{dx^n}\left(e^{-x} x^{n + a}\right)
\eeq
%allows, in the $N \gg1$ limit, to
 one rewrites the second sum in Eq.~(\ref{fNkrec}) as
$\sum_{i=0}^{N}\frac{(k^2/4)^{N+1-i}}{i! (N+1)!} U(-i, 2-i+N, k^2/4)^2$,
which, taking into account (\ref{mu}), yields
\beq
\mu(N) = 2 \int_0^{\infty} v(2\sqrt{x})  \rme^{-2x}
\,\left[L_N(x) L_{N-1}^1(x) - \sum_{i=0}^{N}\frac{x^{N+1-i}}{i! (N+1)!} U(-i, 2-i+N, x)^2\right] \,\rmd x \;,
\label{muN}
\eeq
where $U(a,b,x)$ is the confluent hypergeometric function of the second kind.

Note that  if $v(k)$ is a power, 
\be\label{gen}
v(k) = c\, k^m \;,
\ee
the integration in Eq.~(\ref{EN}) can be performed explicitly, using \cite{PBM}
\be
\int_0^\infty x^{\alpha-1} \rme^{-2x} L_m^\gamma(x) L_n^\lambda(x) \rmd x
= \frac{(1+\gamma)_m\, (1-\alpha+\lambda)_n}{m!\,n!} \, \Gamma(\alpha) \,
{}_3F_2(\alpha,\, \alpha-\lambda,\, 1+\gamma+m;\, \alpha-\lambda-n,\, 1+\gamma;\, -1) \;.
\ee
As a result,
\bea
\mu(N) & = & c\,2^{m+1}\Gamma(\mh+1) \left[
\frac{(2)_{N-1} (-\mh)_{N}}{(N-1)!N!}
{}_3F_2 \left(\mh+1,\, \mh+1,\, N+1;\, \mh-N+1,\, 2;\, -1\right) \right. \\ \nonumber
& & {} + \sum_{i=1}^{N} \left. (-1)^{N-i} \frac{(-\mh+i-N-1)_{N}}{(i-1)!(N+1-i)!}
{}_3F_2 \left(\mh+1,\, \mh+N-i+2,\, N+1;\, \mh-i+2,\, N-i+2;\, -1\right)
\right] 
\label{ENrec}
\eea
(when $v(k)$ is a Laurent series, $\mu(N)$ can be obtained as a
corresponding sum over $m$).

From (\ref{muN}) one can, in the  large $N$ limit,  obtain the asymptotics behavior of $\mu(N)$, at least  in the case 
of  Coulomb interaction $m = -1$.
The first term ${c\over 2\sqrt{x}}  \rme^{-2x}
L_N(x) L_{N-1}^1(x)$   can be simplified   using the asymptotics  
%\beq
%\lim_{n\to \infty} n^{-a} L_n^a (x/n) = x^{-a/2} J_a(2\sqrt{x}),
%\label{bessel}
%\eeq
\beq
\rme^{-x/2} x^{a/2} L_n^a(x) =
\frac{\Gamma(n + a+1)}{(\nu/4)^{a/2} n!} J_a(\sqrt{\nu x}) + O(n^{a/2-3/4})
\eeq
where $J_a (x)$ is the Bessel function, and $\nu = 4 n +2 a +2$.
The second term   yields a sum of integrals which converges  to a constant $  \simeq 0.9$.
As a result,
\beq
\mu(N) \simeq c \left[\sqrt{N} \int_0^{\infty} \rme^{-x}
J_0(\sqrt{(4N+2)x}) J_1(\sqrt{4Nx})\frac{\rmd x}{x} -  0.9\right]
\simeq c \left(\frac{4}{\pi} \sqrt{N} - 0.9\right).
\eeq
The energy  is obtained by integrating
the continuous version of (\ref{mu}), $\rmd E/\rmd N = \mu (N)$
\beq
E(N) \simeq c\left( \frac{8}{3\pi} N^{3/2} -0.9 N\right) \;.
\label{asympcoul}
\eeq

The $N^{3/2}$ scaling is easy to understand.
The number of pairs grows as $N^2$, whereas the characteristic radius
of the system in the ground state, which is the radius of the classical orbit
with $L_\mathrm{max} = N-1$, grows as $\sqrt N$ --- and the same should be true
of the mean interparticle distance.

As an illustration,
we have obtained exact numerical results for $E(N)$ with the Coulomb interaction.
A convenient normalization is $c = \sqrt{2/\pi}$,
which renders the results rational.
Remarkably, exact results can be obtained for up to $N \sim 1000$,
for which a ``brute-force'' calculation, involving $N!$ terms, would
clearly be impossible\footnote{As far as the computation is concerned,
at least in the Coulomb case, using Eq.~(\ref{fNkrec})
and integrating the resulting polynomials turns out to be incomparably faster
than using (\ref{ENrec}). The hypergeometric function
with large $N$ takes much more time to evaluate than a product of two Laguerre
polynomials.}.
Partial results are shown below: 
\begin{center}
\begin{tabular}{|c|l|}
\hline
$N$ & $E(N)$ \\
\hline
2 & $\frac{1}{2}$ \rule{0em}{1.5em} \\
3 & $\frac{87}{64}$ \rule{0em}{1.5em} \\
4 & $\frac{5147}{2048}$ \rule{0em}{1.5em} \\
5 & $\frac{514\,095}{131\,072}$ \rule{0em}{1.5em} \\
10 & $\frac{1\,977\,801\,361\,250\,785}
{140\,737\,488\,355\,328}$ \rule{0em}{1.5em} \\
20 & $\frac{1\,859\,029\,096\,417\,154\,793\,530\,197\,844\,505\,235}
{40\,564\,819\,207\,303\,340\,847\,894\,502\,572\,032}$ \rule{0em}{1.5em} 
%40 & $\frac{1\,913\,080\,915\,096\,003\,822\,940\,524\,331\,993\,889\,094\,992\,365\,381\,082\,903\,884\,839\,128\,295\,321\,815}
%{13\,479\,973\,333\,575\,319\,897\,333\,507\,543\,509\,815\,336\,818\,572\,211\,270\,286\,240\,551\,805\,124\,608}$ \rule{0em}{1.5em} \\
%60 & %$\frac{1\,212\,839\,860\,352\,577\,919\,896\,055\,712\,875\,810\,828\,888\,601\,205\,702\,537\,699\,063\,506\,040\,532\,820\,923\,318\,364\,813\,315\,089\,051\,204\,027\,134\,565\,755}
%{4\,479\,489\,484\,355\,608\,421\,114\,884\,561\,136\,888\,556\,243\,290\,994\,469\,299\,069\,799\,978\,201\,927\,583\,742\,360\,
%321\,890\,761\,754\,986\,543\,214\,231\,552}$ 
\rule{0em}{1.5em} \rule[-0.7em]{0em}{1em} \\
\hline
\end{tabular}
\end{center}

\begin{figure}
\begin{center}
\includegraphics[scale=1]{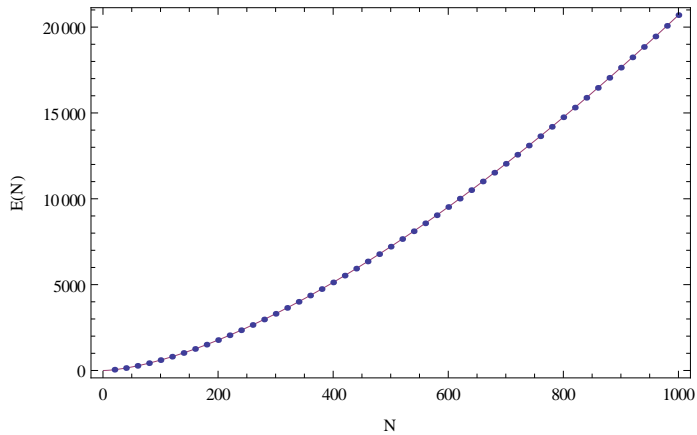}\label{fig1}
\caption{The function $E(N)$ (dots) versus Eq. (\ref{asympcoul}) (continuous curve).}
\end{center}
\end{figure}

A plot of the (discretized) logarithmic derivative,
\be
D(N) = \frac{\log E(N) - \log E(N-1)}{\log N - \log (N-1)} \;,
\ee
in Fig.~2 is clearly consistent with
$\lim_{N \to \infty} D(N) = \frac{3}{2}$, in accordance with Eq. (\ref{asympcoul}).

\begin{figure}
\begin{center}
\includegraphics[scale=1]{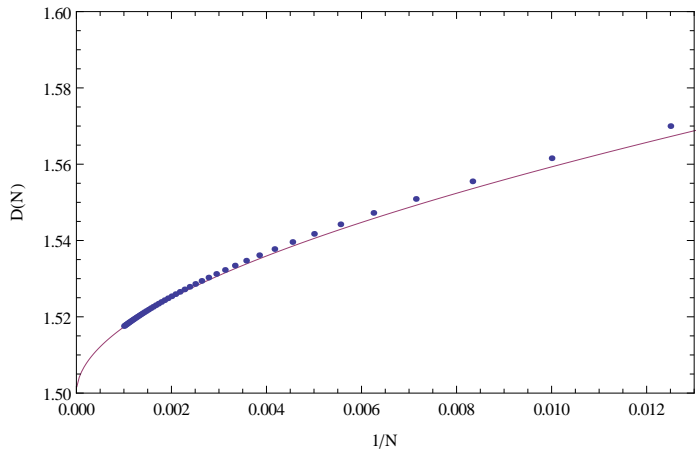}\label{fig2}
\caption{The dicretized logarithmic derivative $D(N)$ as a function of $1/N$ (dots) versus
$\partial E(N)/\partial \log N$,
with $E(N)$ given by Eq. (\ref{asympcoul}) (continuous curve).}
\end{center}
\end{figure}

\section{The three-body Coulomb case}

Coming back to Eq.~(\ref{pV}), for Coulomb interaction, one has
\beqa
P \sum_{i<j=1}^N V(|z_i - z_j|) \chi & = & c \sum_{i<j=1}^N\sum_{n_1, n_2 = 0}^\infty
\frac{(-1)^{n_1}\Gamma (n_1 + n_2 +\frac{1}{2})}{n_1!\,n_2!\,(n_1+n_2)!}
\frac{(z_i -z_j)^{n_1+n_2}}{2^{n_1+n_2+\frac12}}\frac{\D^{n_1+n_2}}{\D z_i^{n_1}\D z_j^{n_2}}
\chi (z_1, ...,z_N) \;\nonumber \\
{} & = & c \sum_{i<j=1}^N\sum_{n= 0}^\infty
\frac{\Gamma (n +\frac{1}{2})}{(n!)^2 2^n}(z_i -z_j)^n (\D_j -\D_i)^n
\chi (z_1, ...,z_N) \;.
\label{pVC}
\eeqa
This can be directly diagonalized in each sector with a given
number of particles $N$ and angular momentum $L$,
with the basis formed by (anti)symmetrized functions of the form (\ref{basis}).
The general structure of the spectrum is similar to the delta interaction case:
There are center-of-mass excitations, so that above each state
with energy $E$ there is a ``tower'' with energies $E+n$, $n = 1, 2, \ldots$~.
Only the ``pure relative'' eigenstates, devoid of these excitations, are of interest.
We restrict ourselves to the 3-body problem.
For the bosons with delta interaction, all the 3-body states
turned out to have rational energies \cite{LLL} (with a suitable choice
of an overall factor $c$).
With Coulomb interaction, though, irrational values start appearing
rather low in the spectrum, for bosons and fermions like.
Scaling away the overall irrationality, as before, by putting $c = \sqrt{2/\pi}$,
one finds all the eigenvalues of ``pure relative'' states,
up to the appearance of irrationalities, are, for bosons:
\begin{center}
\begin{tabular}{|c|l|}
\hline
$L$ & $E$ \\
\hline
0 & 3 \\
2 & $\frac{33}{16}$ \rule{0em}{1.5em} \\
3 & $\frac{51}{32}$ \rule{0em}{1.5em} \\
4 & $\frac{1755}{1024}$ \rule{0em}{1.5em} \\
5 & $\frac{3153}{2048}$ \rule{0em}{1.5em} \\
6 & $\frac{3(27749 \pm 7\sqrt{766249})}{65536}$ \rule{0em}{1.5em} \rule[-0.7em]{0em}{1em}
%$\frac{3(27749 + 7\sqrt{766249})}{65536}$
\\
\hline
\end{tabular}
\end{center}
and for fermions:
\begin{center}
\begin{tabular}{|c|l|}
\hline
$L$ & $E$ \\
\hline
3 & $\frac{87}{64}$ \\
5 & $\frac{4881}{4096}$ \rule{0em}{1.5em} \\
6 & $\frac{4119}{4096}$ \rule{0em}{1.5em} \\
7 & $\frac{140283}{131072}$ \rule{0em}{1.5em} \\
8 & $\frac{63255}{65536}$ \rule{0em}{1.5em} \\
9 & $\frac{3 (10025047 \pm 3 \sqrt{107141413705})}{33554432}$ \rule{0em}{1.5em} \rule[-0.7em]{0em}{1em}
%$\frac{3 (10025047+3 \sqrt{107141413705})}{33554432}$
\\
\hline
\end{tabular}
\end{center}
%The systematics of the rational eigenvalues as well as the
%structure of the eigenfunctions have yet to be understood.

\section{DISCUSSION}
The opportunity to  calculate the energy of an eigenstate
of $N$ interacting two-dimensional bosons or fermions is certainly
due to the fact that the LLL projection simplifies the situation.
It reduces the dimension of the single-particle phase space  \cite{1D}
%(the effective problem becomes one-dimensional );
and even more importantly, the ``ground state''
(Bose condensate for bosons, Laughlin state for fermions) ends up being an eigenstate
of the interacting Hamiltonian, which means that all one has to compute
for that state is a single matrix element.
Nevertheless, even this simplified setup has a physical meaning,
which makes our results applicable to real systems.

The relevant case is when the LLL is separated by a gap
from the rest of the spectrum. This happens
when the whole system rotates with angular speed $\omega$,
or if there is a strong magnetic field.
% instead of the harmonic potential.
But if the LLL is flat (which happens if there is a magnetic
field but no harmonic potential), all the  LLL $N$-body states
have the same degenerate energy. Our result for the ground'' state,
with the minimum $L$, is valid (it still does not mix with the other states),
but not meaningful physically, as that state is not separated by an energy gap
from states with higher values of $L$.
This changes if  a harmonic potential adds
$L(\omega_t-\omega_\mathrm{c})$ to the energy.
One can then claim that if the interaction is weak enough compared to the gap,
the exact energy of the $N$-body ground state is known.

One has to be careful, however, when taking the thermodynamic limit.
As soon as $N(\omega_t-\omega_\mathrm{c})$ becomes bigger than $\omega_\mathrm{c}$, the energy of
the lowest single-particle state in the first LL becomes smaller than that
of the $N$-th single-particle state in the LLL.
Actually, the LLL projection approximation breaks
as soon as $N(\omega_t-\omega_\mathrm{c}) \sim \omega_\mathrm{c}$.
Therefore, for our result to be interesting, the $\omega_\mathrm{c} \to \infty$ limit
has to be taken first, and then the thermodynamic limit $N \to \infty$.
%(This is of course the same situation as with LLL anyons.)

Finally the same techniques could be applied to excited states
with higher values of $L$. To do so one woud have to evaluate the matrix element
$\langle p'_1 p'_2 \ldots p'_N | P\sum_{i<j=1}^N V(|z_i - z_j|) | p_1 p_2 \ldots p_N \rangle$,
where $| p_1 p_2 \ldots p_N \rangle = \prod_{l=1}^N z_l^{p_l}$, properly symmetrized
or $| p_1 p_2 \ldots p_N \rangle = \prod_{i<j=1}^N (z_i - z_j)\prod_{l=1}^N z_l^{p_l}$ properly antisymmetrized.

\end{document}